\documentclass{article_arxiv}

\makeatletter
\renewcommand*{\@seccntformat}[1]{\csname the#1\endcsname\hspace{0.1cm}}
\makeatother

\usepackage{float}
\usepackage{amssymb,amsmath}
\usepackage{graphicx}

\begin{document}
\title{\vspace{-2.5cm} \bf \Large Optimal protocol for a collective flashing ratchet.}
\author{ F. Roca$^{1}$, J. P. G. Villaluenga$^{1}$, L. Dinis$^{2,3}$ }

\maketitle
\vspace{-1cm}
\begin{center}
$^{1}$  Departamento de F\'isica Aplicada I,  Universidad Complutense de Madrid - 28040 Madrid, Spain\\
$^{2}$ 	Departamento de F\'isica At\'omica, Molecular y Nuclear, Universidad Complutense de Madrid - 28040 Madrid, Spain\\
$^{3}$ GISC -- Grupo Interdisciplinar de Sistemas Complejos, Spain
\end{center}

\begin{center}
\parbox{0.85\linewidth}
{We study a system of independent Brownian particles in a flashing ratchet potential that can be turned on or off depending on the position of the particles, with the aim of maximising the speed of the center of mass in the long run. First, an explanation on how to find the optimal protocol {using Bellman's principle of optimality} for any number of particles is given. Then the problem is numerically solved for a system of 2 particles. Simulations show that the optimal protocol performs better than the maximisation of instantaneous center-of-mass speed, a protocol known to give better results for 2 particles than any open-loop protocol.}
\end{center}
{\sc pacs} numbers: 05.40.-a, 02.30.Yy

\section{ Introduction}
Control in Brownian ratchets has been attracting considerable attention in recent years mainly due to their potential applications in diverse fields \cite{Roman, Reimann, Savel, Linke}. There are two control strategies: open-loop and closed-loop. In the first kind of strategies, the tuning of the control parameter is made without using any information about the state of the system. In contrast, in the second control policy, the modulation of the control parameter depends on the state of the system. In the present work, a close-loop control protocol is implemented in a system where the particle positions are monitored. 

The system consists on an ensemble of Brownian particles in a flashing ratchet potential that can be switched on or off depending on the position of particles, with the aim of maximising the center-of-mass position of the system. Previous papers have shown that some closed-loop protocols are optimal for one particle and perform better than any periodic open-loop flashing for
ensembles of moderate sizes, but is defeated by random or periodic switching for large ensembles \cite{Dinis,Cao}.

Nevertheless, the search of an optimal protocol to maximise the flux of particles is still an open question \cite{FeitoCao}. In this letter we propose a feedback control protocol built by using tools of the theory of control in stochastic systems, in particular, the principle of Bellman. Due to the generality of the principles employed, the results are useful to optimise systems of small size in which the fluctuations play a dominant role. 

\section{The model}
	The system we consider here is an ensemble of $N$ overdamped Brownian particles at temperature $T$ in an asymmetric periodic potential that can be swiched on or off. The dynamics of the system is described by the Langevin equation

\begin{equation} \label{langevin}
\eta \dot{x_{i}}(t)=\alpha(t) F(x_{i}(t))+\xi_{i}(t);\qquad i=1,\cdots, N,
\end{equation}

where $x_i(t)$ is the position of particle $i$, $\eta$ is the friction coefficient and $\xi_i (t)$ are the thermal noise with zero mean and correlation $\langle \xi_{i}(t) \xi_{j}(t')\rangle = 2\eta kT \delta_{ij}\delta(t-t')$. The parameter $\alpha(t)$ is the control parameter which is assumed that can take on values 1 and 0 depending whether the potential is switched on or off, respectively. The force, which represents the effect of the potential when the ratchet is on, is given by ${F(x)=-V'(x)}$. We have considered as the ratchet potential ${V(x)}$ \cite{Reimann}:
	
\begin{equation}	\label{potential}
V(x)=V_{0} \Big( \sin(\frac{2 \pi x}{L})+0.25\sin(\frac{4\pi x}{L}) \Big)
\end{equation}

with ${L=3}$. Without loss of generality, we can set $\eta=1$ and $kT=1$, giving a diffusion coefficient $D=kT/\eta=1$ and leaving potential intensity $V_0$ as the only free parameter.

The Langevin equation can be written for each particle using the forward Euler discretisation as follows:
	
\begin{equation}	
\label{euler}
\begin{split}
x_i(t_k)= x_i(t_{k-1}) +\alpha_k F(x_i(t_{k-1})) \Delta t + \Delta W; \\
 i=1,\ldots, N; k=0,\ldots,M; t_k=k\Delta t, \alpha_k=\alpha(t_k).
\end{split}
\end{equation}

The increment ${\Delta W}$ represents the thermal noise. We have used a normally distributed  pseudo-random number with mean 0 and standard deviation $\sigma=\sqrt{2D\Delta t} $. The time step used in the simulations is set to ${\Delta t}=0.005$. 

We will be interested in maximising the average particle flow after some, generally large, operation time $t_M$
\begin{equation}
\langle \bar x(t_M)\rangle=\frac{1}{N}\sum_{i=1}^{N}\langle x_i(t_M)\rangle.
\end{equation}
Here and in the following $\bar\cdot$ represents a particle average and $\langle \cdot\rangle$ an average over realizations of the thermal noise.
The average particle flow can be written in terms of the position change in every step $\Delta x_i(t_k)=x_i(t_k)-x_i(t_{k-1})$ as follows:

\begin{equation}	\label{eq.4}
\langle \bar x(t_M) \rangle=\frac{1}{N}\sum_{i=1}^N\sum_{k=1}^{M} \langle \Delta x_i(t_{k}) \rangle= \sum_{k=0}^{M-1} {\langle \alpha_{k} \bar F(t_{k})\rangle} \Delta t
\end{equation}

where we have used the zero mean property of the noise and the force per particle $\bar F(t_{k})=\frac{1}{N}\sum_{i=1}^N F(x_i(t_k))$.

As the time step $\Delta t$ is constant, the maximisation of the particle flow is equivalent to the maximisation of:

\begin{equation}	\label{eq_JM}
	J_{M-1}(\vec x,t_{M-1}{,t_{M-2},\dots,t_0} )\equiv \sum_{k=0}^{M-1} {\langle \alpha_k  \bar F(t_k) \rangle}
\end{equation}	
with $\vec x=(x_1,x_2,\ldots,x_N)$. {$J_{M-1}$ represents the objective function to maximise with respect to the choice of the state of the potential $\alpha$ as a function of the position of the particles and the number of decisions still to make in the trajectory.}

\begin{figure}[h]
  \centering
\includegraphics[width=0.6\textwidth]{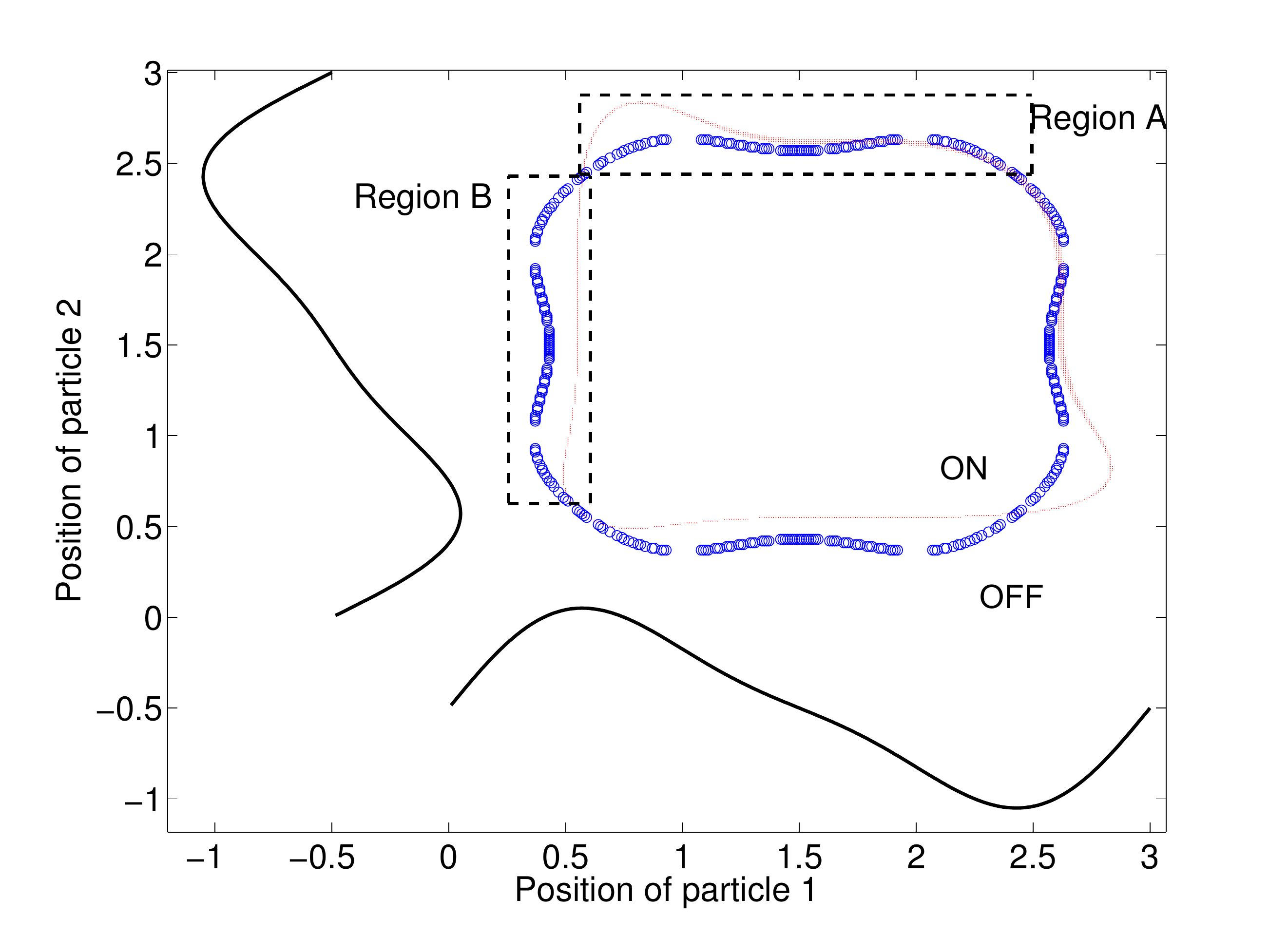}
\caption{Critical boundaries for $D=1$, $V_0=5kT$, one step (blue circles) and 100 steps before the end of operation (steady state) (red thin line). The black solid line represents one period of the potential felt by the particles.\label{fig.1} }
\end{figure}

{The procedure to maximise functions $J_{M-1}$ is backwards recursive, so we will start by optimising a trajectory of just one time step, with only one decision to make.}
Rewriting Eq.~\eqref{eq_JM} for just one step ($M=1$) and optimising we obtain:
	
\begin{equation}	\label{eq.local}
\hat{J}_0(\vec{x}(t_{0}))= \left\{\begin{array}{ll} 
\bar F(t_{0}), &  \textrm{if } \bar F(t_{0}) >0, {(\alpha_0=1)}  \\ 0, &   \textrm{if }\bar F(t_{0})<0 , {(\alpha_0=0)}
\end{array} \right.
\end{equation}
{Inside parenthesis we have indicated the value of $\alpha_0$ to be chosen to obtain the maximum average forward movement $\hat J_0$ in each case.}
 From condition \eqref{eq.local}, in principle, we can obtain a region in a {$N$}-dimensional space in which the potential should be on {and a region where it should be off}. It is depicted in figures \ref{fig.1} and \ref{fig.2} for the case of 2 particles {by means of the boundary separating both regions. }
 


	In \cite{Cao,Dinis} a protocol that maximises instantaneous center-of-mass speed is analysed and performs better than any periodic or random switching for low number of particles. This {\em local} time protocol consists in switching the ratchet potential on when $\bar F >0 $ and switching it off when $\bar F <0 $, 
 {which corresponds to using the criterion for $\alpha$ expressed in Eq.~\eqref{eq.local} at every time step in the trajectory. We will compare the performance of the global optimal decision protocol against this local one.}

	To find a {\em global} time strategy, one that optimises the total center-of-mass motion after a number of steps, we have applied Bellman's criterion\cite{Layton}. Bellman's criterion allows to find the complete sequence of optimal decisions by backwards recursion. The optimal strategy can be calculated in terms of the state $\vec x(t_k)$ of the system in each time step, which allows the application of the criterion as a feedback protocol. Bellman's criterion establishes that given a certain initial condition (in this case, a certain position for each particle) the optimal sequence of decisions is independent of whichever the past strategies that led to the given initial condition. This breaks a dynamics optimization problem into simpler subproblems.
	
	{One of the  consequences of the Bellman's criterion is that Eq. \eqref{eq.local}  constitutes an {\em optimal decision map} that can be used for the last step of every trajectory to ensure maximum displacement.}

\begin{figure}[h]
  \centering
\includegraphics[width=0.49\textwidth]{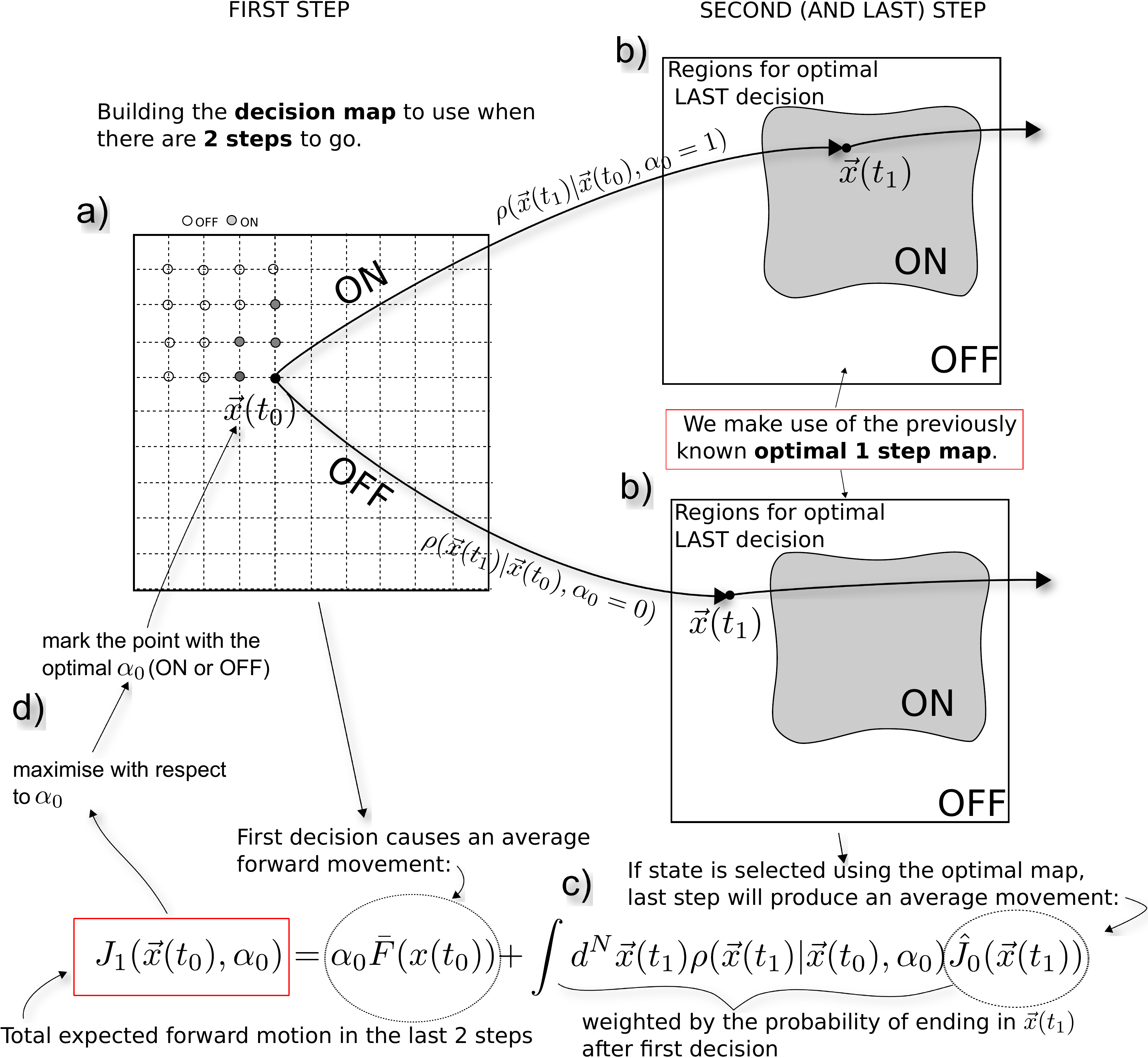}
\caption{ {Schematic of the two-step decision process for $2$ particles described in the text. a) In the first step, both possible decisions are tried at a given $\vec x(t_0)$. An advancement $\alpha_0\bar F(x(t_0))$ is produced. b) After evolution, the state is now $\vec x(t_1)$. Then, the previous optimal decision map is used to ensure maximum advancement $\hat J_0(\vec x(t_1))$ in the last step. c) This forward motion has to be weighted by the probability of getting from $\vec x(t_0)$ to $\vec x(t_1)$ with a given state of the potential $\alpha_0$. d) The total average forward motion $J_1(\vec x(t_0),\alpha_0)$ is then maximized with respect to $\alpha_0$ and the optimal decision assigned to the point $\vec x(t_0)$. The process is then repeated for the next $\vec x(t_0)$ in the grid in order to build the full decision map for a two-step process.\label{fig_schematic_2}} }
\end{figure}

Let us now imagine we have a 2-step trajectory to optimise and we start in a state of the system $\vec x(t_0)$. {The schematic of the following decision process is depicted in Fig.~\ref{fig_schematic_2} for the case of 2 particles (for reasons that will become apparent later, we have discretized the full space of positions in a two-dimensional grid).} In principle, we can choose to switch on or off the potential at this first time step.  After this first decision, particles will jump stochastically and end up in a different state $\vec x(t_1)$. Bellman's criterion tells us that whatever decision we took before, the optimal decision protocol should now choose according to \eqref{eq.local}, as this is the last step. {Thus, Bellman's criterion allows us to fix the optimal choice for the last step so that the value of $J_1 (\vec x(t_0),\alpha_0)$ and the optimal decision $\alpha_0$ depend only on the position of the particles at $t_0$ and not at $t_1$. This allows us to construct an optimal decision map to use when there are still two steps before the end of the trajectory.}

In the final step, an advancement of $\hat J_0(\vec x(t_1))$ will be obtained, but only with a probability $\rho(\vec x(t_1)|\vec x(t_0),\alpha_0)$ of jumping from $\vec x(t_0)$ to $\vec x(t_1)$ with  $\alpha_0$ state of the potential. Averaging, the total movement in these 2 steps would be:

\begin{equation}
\label{eq_J1J0}
\begin{split}
J_1(\vec x(t_0),\alpha_0)=\alpha_0\bar F(x(t_0)) \\+ \int_{-\infty}^{\infty}d^N\vec x(t_1) \rho(\vec x(t_1)|\vec x(t_0),\alpha_0) \hat J_0(x(t_1))
\end{split}
\end{equation}

{We could now compare the result of $J_1$ for  $\alpha_0=\{0,1\}$
 to obtain the optimal decision for $\alpha$ at $t_0$ (the best first decision),} and the corresponding maximum forward movement for 2 time steps
\begin{equation}
\hat J_1(\vec x(t_0))=\max\{J_1(\vec x(t_0),\alpha_0=0),J_1(\vec x(t_0),\alpha_0=1)\},
\end{equation}
as a function of $\vec x(t_0)$. 

To proceed further, we need to compute the probability density $\rho(\vec x(t_k)|\vec x(t_{k-1}),\alpha(t_{k-1}))$ of a general jump. Using Langevin equation (\ref{langevin}) we deduce than in order for the particle to end in $\vec x(t_k)$ after a $\Delta t$ starting from $\vec x(t_{k-1})$, the Brownian increments have to fulfill:

\begin{equation}        \label{eq.9}
{\Delta W_i}= \left[x_i(t_k) - x_i(t_{k-1})\right]-\alpha_{k-1}{F(x_i(t_{k-1}))} \Delta t  
\end{equation}
 
The quantities $\Delta W_i$ are normally distributed and uncorrelated with each other and thus:
	
\begin{equation}  
\begin{split}
\rho( \vec{x}(t_{k})|\vec{x}(t_{k-1}),\alpha_{k-1})=\frac{1}{(2 \pi 2 D \Delta t)^{\frac{N}{2}}} \times & \\ \exp{\left(-\frac{\sum_j^N\big(x_{j}(t_k) - x_{j}(t_{k-1})-\alpha_{k-1} F(x_{j}(t_{k-1}))\Delta t \big)^2}{2D \Delta t}\right)}&
\end{split}
\label{eq.10}	
\end{equation} 

Finally we just have to substitute \eqref{eq.10} in \eqref{eq_J1J0} to compute the needed integrals.

Finding the optimal decisions for a longer trajectory requires a completely analogous recursive method. Applying the recurrence to the last $k$ steps of a trajectory that starts at $\vec x(t_{M-k})$, the optimal $k$-th decision can be found from the knowledge of $\hat J_{k-1}$ by comparing $J_{k}(\vec x(t_{M-k}),0)$ and $J_{k}(\vec x(t_{M-k}),1)$, with:

\begin{equation}
\label{eq_JkJk-1}
\begin{split}
J_k&(\vec x(t_{M-k}),\alpha_{M-k})=\alpha_{M-k}\bar F(\vec x(t_{M-k}))\\ +& \int_{-\infty}^{\infty}d^N\vec x(t_{M-k+1}) \rho(\vec x(t_{M-k+1})|\vec x(t_{M-k}),\alpha_{M-k}) \\\hat J_{k-1}&(\vec x(t_{M-k+1}))
\end{split}
\end{equation}

Solving the recursion analytically for any number of steps resulted a challenging task. Instead one can discretise the space of particle positions and solve the integrals numerically at every point of an $N$-dimensional grid.  
We were able to compute $J_k$ functions for the case of two Brownian particles, where we can already notice a substantial difference between the global and local optimisation and which simplifies both the graphical representation and the identification of the mechanisms involved in explaining the differences. 

For two particles, it suffices to compute the value of every $\hat J_k$ at every point of a grid in the square $(x_1,x_2)\in[0,L)\times[0,L)$ with periodic boundary conditions, due to the periodicity of the potential.
The integrals were calculated in an interval centered in every point of the grid $(x_1(t_{M-k}),x_2(t_{M-k}))$ with three standard deviation of length in each direction and using the boundary conditions.

\section{Results}
       
	The result of the calculations returns for each point of the square grid both the optimal decision to take with regard to switching or not the potential and the maximum average center of mass movement corresponding to that (and subsequent) optimal decision. This optimal maps (one for each time step) consist of two distinct regions separated by a critical boundary where either decision yields the same results on average.
If the position of the particles corresponds to a point in the region inside the border, the optimal choice is switching on the potential; whereas if the position corresponds to a point outside the border, the optimal choice is switching it off.  For the sake of clarity we will just show the critical boundary in the figures.

        When Bellman's criterion is applied backwards in time recursively, the border experiences deformations step by step.  After a sufficiently large number of steps, a stationary state is achieved and the critical boundary does not vary between a given instant and the previous one. Actually, the stationary critical region is the most relevant one because it gives the optimal decision policy when operating the ratchet in the steady state.
        
         In fig.\ref{fig.1}, we show the critical boundary of the system for the last step and for 100 steps before, when the system is in a steady state, for a diffusion coefficient of $D=1$ and $V_0=5kT$. A representation of the potential is provided as well in order to facilitate data reading. As explained before, the map for the last step corresponds to the choices the local time protocol will follow at every instant.

\begin{figure}[h]
  \centering
\includegraphics[width=0.6\textwidth]{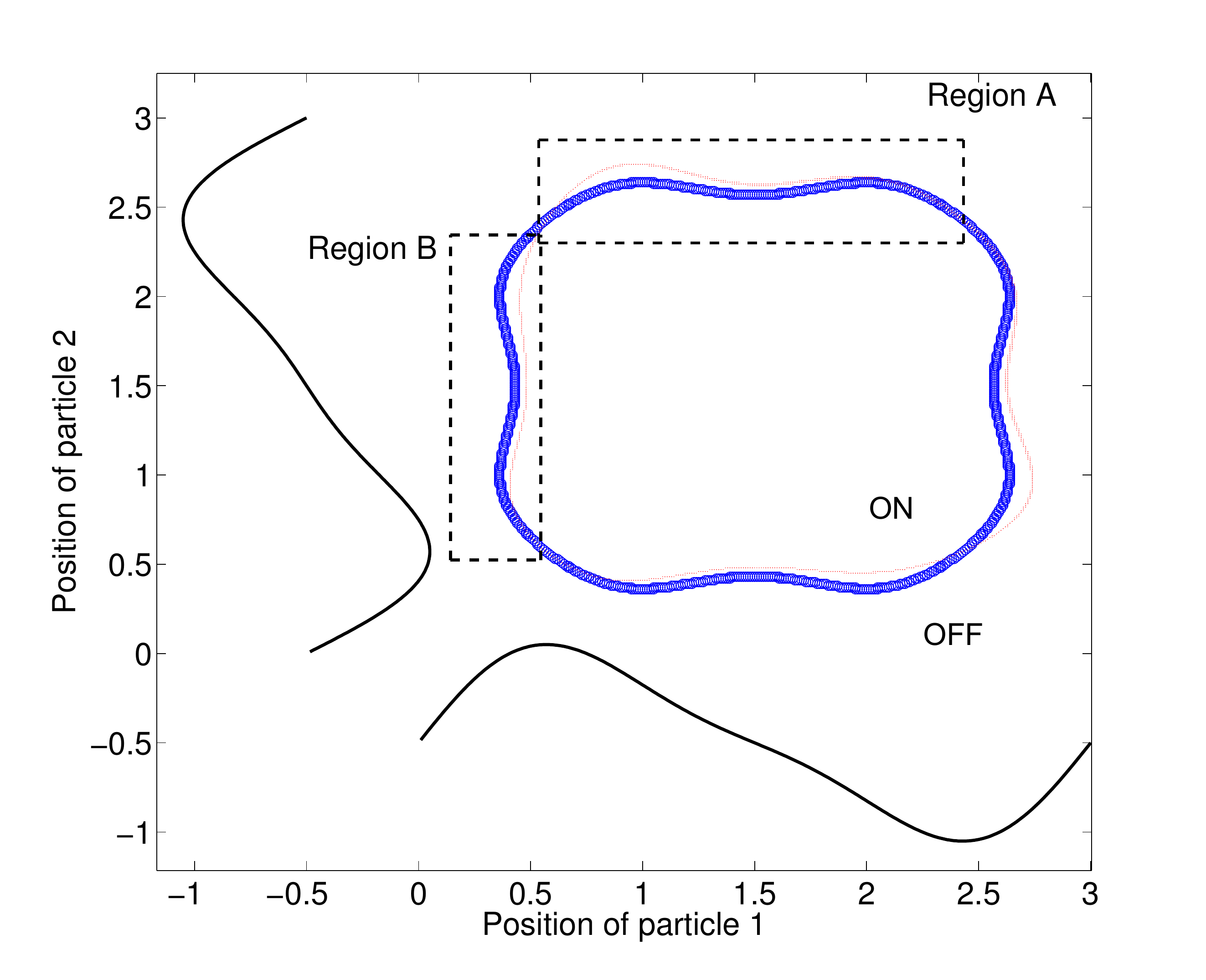}
\caption{Critical boundaries for $D=1$, $V_0=1kT$, one step (blue thick line) and 100 steps before the end of operation (steady state) (red thin line).\label{fig.2} The black solid line represents one period of the potential felt by the particles. }
\end{figure}

        We carried out the same computations for a lower value of the potential. Fig.~\ref{fig.2} represents the critical boundary for $D=1$, $V_0=1kT$.

\begin{figure}[h]
  \centering
\includegraphics[width=0.6\textwidth]{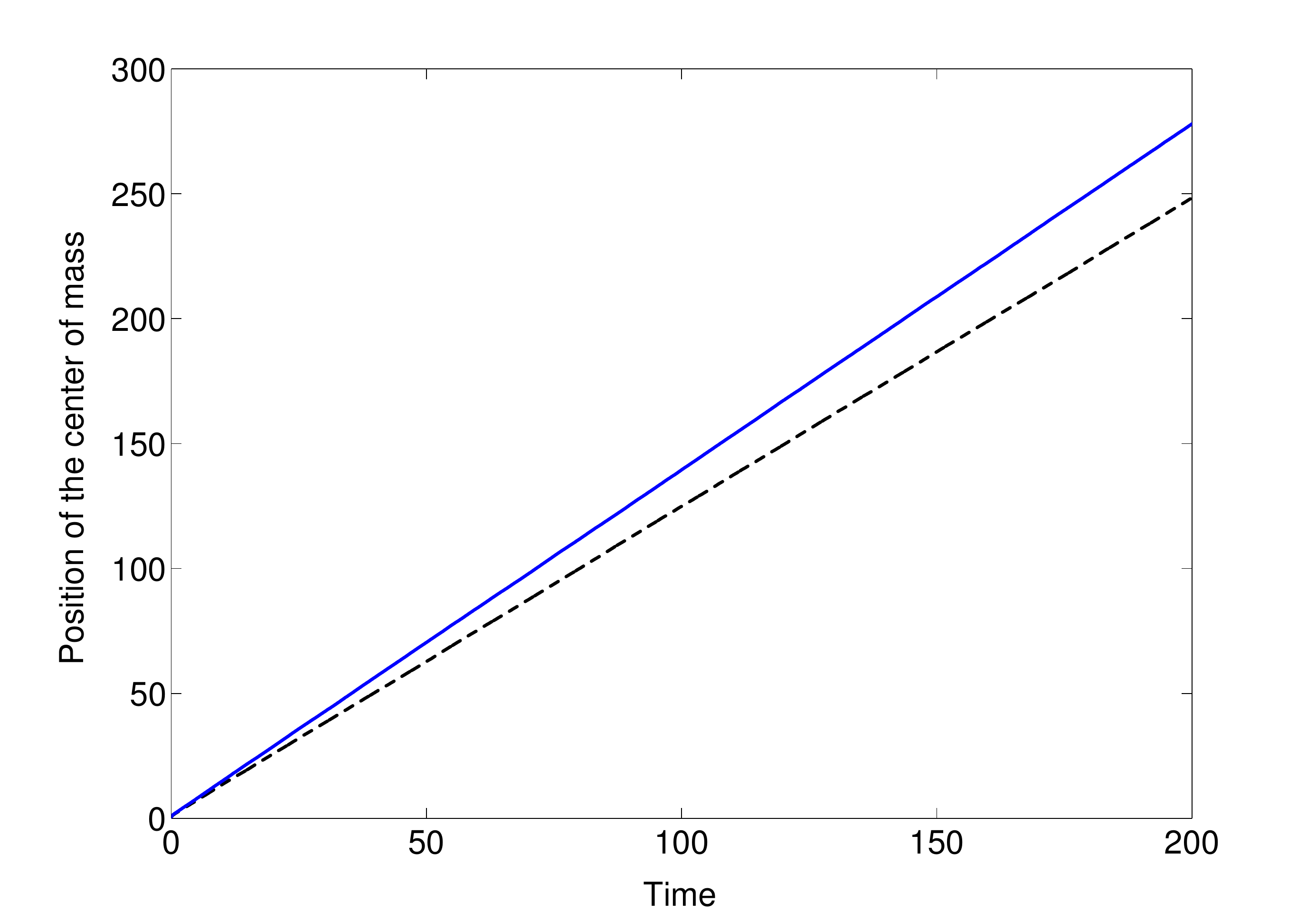}
\caption{Langevin equation simulation using the local time protocol (black dashed line) and the stationary global strategy (blue solid line) for $D=1$, $V_0=5kT$.}
\label{fig.3}
\end{figure}

Using Langevin equation \eqref{langevin}, we have simulated  the motion of two particles when operating following the prescription determined through Bellman's criterion and compared it to the result when the local time protocol is used. The results are plotted in fig.~\ref{fig.3}.
        
        Each point in the figure is the mean of the position of the center of mass over 100 simulations. It is clear that Bellman's strategy (blue) produces a larger center of mass shifting than the local time strategy (black).
        
For a lower value of the potential ($V_0=1kT$), both strategies become very similar and the global one only beats the local one by a smaller amount (not shown).
        

\section{Discussion}
Following the prescription computed using Bellman's criterion produces a higher center-of-mass speed than local optimisation in the long run. Figures \ref{fig.1} and \ref{fig.2} are suggestive of a possible explanation for this effect. There are essentially two regions where the global and local strategies differ, marked in figure \ref{fig.1} with dashed rectangles for clarity. The other two unmarked regions are simply the reflections of the latter by the bisector of the first quadrant, which correspond to switching particle positions and therefore the same explanation applies, switching particle labels 1 and 2.

Region A corresponds to a situation where particle 1 is near or not too far from the maximum and feels a positive force and particle 2 is in a negative force region but close to the minimum of the potential. In this situation, local optimization switches off the potential since the total force on the particles is negative due to the asymmetry of the potential. It is however not difficult to predict what will happen if we nevertheless switch on the potential and let the system evolve as the global optimization strategy suggests, at least in the limit of low noise. Initially the center of mass will in fact go backwards since total force is negative (particle 2 will move backwards faster than particle 1 will move forward). However, after a short time particle 2 will reach the minimum of the potential and effectively stop on average (it will fluctuate around the minimum). In the meantime, particle 1 will probably move forward until it reaches the next minimum, compensating and even exceeding particle 2 retrograde movement. As a result, the center of mass will have moved forward when particle 1 reaches the next minimum. Global optimization beats local optimization by using a prediction of what will happen based on the probability of each possible evolution computed from the dynamics.

Let us turn our attention now to region B, where particle 1 is close to the maximum in a region of small negative force and particle 2 in a positive force region of the potential. Local optimization prescribes switching it on. The most probable outcome of this action is that particle $2$ will move forward and particle 1 will move backwards away from the maximum until the backward force is big enough and the local protocol prescribes switching the potential off. In order for the center of mass to continue moving forward, particle 1 must diffuse with the potential off until it reaches again the position of the maximum and overcomes it. 
On the other hand, the global protocol prescribes keeping the potential off 
 and wait until it crosses the position of the maximum of the potential. Compared to the situation described for the local protocol, now the distance to overcome by diffusion is smaller, and so is the time on average one has to wait with the potential off.

Combining these two mechanisms the global protocol beats the local one, which already produces a faster movement for 2 particles compared to the open-loop protocols.

Increasing the noise-potential ratio in the system, that is, increasing temperature or conversely decreasing the potential, modifies the global stationary map. As explained before, the global criterion relies on a prediction of what will happen based on the dynamics of the system. As the evolution becomes noisier, future positions of the particles become increasingly uncorrelated to actual positions, and the global protocol necessarily becomes more local in time, its boundary becoming closer to the boundary for the local protocol. 

\section{Conclusions}

We have devised a close-loop control protocol for a system of independent Brownian particles in a flashing ratchet potential. The aim of the protocol, which is based on a switching policy depending on the position of the particles, is the maximisation of the center of mass of the system. The strategy is based on the Bellman's criterion that allows to find a complete sequence of optimal decisions by a backwards recursive method. It is worth noting that this method is a global optimisation protocol, in contrast to usual control strategies which are local. The optimal protocol was solved numerically for a systems of two Brownian particles, and it produces a larger shifting of the center of mass, in the steady state, than the local protocol. The algorithm is very general and could be applied to any system with white noise (either Gaussian or not) and to different target functions, not just the average current.

Finally, the optimal closed-loop strategy could be implemented in systems where the particles are monitored, as in some experimental realizations of colloidal particles in liquids \cite{Marquet, Bader}
\section{Acknowledgments}
L.D.~ acknowledges support from grant ENFASIS (FIS2011-22644) form Spanish Government and MODELICO-CM (S2009/ESP-1691) from Comunidad de Madrid, Spain. F.R.~ acknowledges financial support from Ministerio de Educaci\'on, Cultura y Deporte (Spain) through Becas de Colaboraci\'on program.

\end{document}